\documentclass[twocolumn,prl, twocolumn,superscriptaddress, nofootinbib]{revtex4-1}

\usepackage[T1]{fontenc}
\usepackage[latin9]{inputenc}
\usepackage{color}
\usepackage{amsmath}
\usepackage{amssymb}
\usepackage{graphicx}
\usepackage{wasysym}
\usepackage[unicode=true,pdfusetitle,
 bookmarks=true,bookmarksnumbered=false,bookmarksopen=false,
 breaklinks=false,pdfborder={0 0 0},pdfborderstyle={},backref=false,colorlinks=true]
 {hyperref}
\hypersetup{
 citecolor=blue,urlcolor=blue}

\makeatletter

\newcommand{\noun}[1]{\textsc{#1}}

\usepackage{bbm}

\makeatother

\begin{document}
\title{The hierarchy recurrences in local relaxation }
\author{Sheng-Wen Li }
\affiliation{Center for quantum technology research, School of Physics, Beijing
Institute of Technology, Beijing 100081, China }
\author{C. P. Sun }
\affiliation{Graduate School of China Academy of Engineering Physics, Beijing 100193,
China}
\affiliation{Beijing Computational Science Research Center, Beijing 100193, China}
\begin{abstract}
Inside a closed many-body system undergoing the unitary evolution,
a small partition of the whole system exhibits a local relaxation.
If the total degrees of freedom of the whole system is a large but
finite number, such a local relaxation would come across a recurrence
after a certain time, namely, the dynamics of the local system suddenly
appear random after a well-ordered oscillatory decay process. It is
found in this paper, for a collection of \noun{n} two-level systems
(TLSs), the local relaxation of one TLS within has a hierarchy structure
hiding in the randomness after such a recurrence: similar recurrences
appear in a periodical way, and the later recurrence brings in stronger
randomness than the previous one. Both analytical and numerical results
that we obtained well explains such hierarchy recurrences: the population
of the local TLS (as an open system) diffuses out and regathers back
periodically due the finite-size effect of the bath {[}the remaining
$(\text{\textsc{n}}-1)$ TLSs{]}. We also find that the total correlation
entropy, which sums up the entropy of all the \noun{n} TLSs, approximately
exhibit a monotonic increase; in contrast, the entropy of each single
TLS increases and decreases from time to time, and the entropy of
the whole \noun{n}-body system keeps constant during the unitary evolution. 
\end{abstract}
\maketitle
\noindent\textbf{Introduction: }When an open system is contacted
with a bath infinitely large, the open system would approach a certain
steady state after a long time relaxation. However, such an irreversible
behavior cannot be seen in the dynamics of one or few-body systems.
Thus the macroscopic irreversibility seems contradicted with the microscopic
reversibility \citep{hobson_irreversibility_1966,prigogine_time_1978,mackey_dynamic_1989,uffink_compendium_2006,swendsen_explaining_2008}. 

One useful way to look through this problem is to study the system
relaxation contacted with a finite bath, namely, the bath contains
a finite number of degrees of freedom (DoF), and then consider its
transition to the thermodynamics limit \citep{cramer_exact_2008,flesch_probing_2008,eisert_quantum_2015,xu_noncanonical_2014}.
The open system and the bath as a whole isolate system always follows
the unitary evolution and keeps a constant entropy as the initial
state, while the open system itself seems relaxing towards a certain
steady state, thus such a relaxation behavior of the open system itself
is called the \emph{local relaxation} \citep{cramer_exact_2008,eisert_quantum_2015,flesch_probing_2008}.

Due to the finite-size effect of the bath, the local relaxation of
the open system would come across a recurrence after a certain time:
at first the system dynamics shows a well-ordered oscillatory decay
behavior, but then suddenly becomes random \citep{hanggi_reaction-rate_1990,zwanzig_nonequilibrium_2001,cramer_exact_2008,flesch_probing_2008,eisert_quantum_2015}.
With the increase of the DoF number in the bath, such a recurrence
time appears much later, thus it does not show up in practice. 

In this paper, we find that, in the region after such a recurrence,
indeed there exists a hierarchy structure hiding in the randomness:
similar recurrences appear in a periodical way, and the later recurrence
brings in stronger randomness than the previous one, therefore, we
call them \emph{hierarchy recurrences}. 

Here we study the dynamics of a chain of \noun{n} two-level systems
(TLSs). One of the TLSs is treated as the open system, and all the
other $(\text{\textsc{n}}-1)$ TLSs make up a finite bath. We obtain
a Bessel function expansion for the system dynamics, which well explains
the appearance of such hierarchy recurrences. Further, we also find
the physical reason for the appearance of such hierarchy recurrences:
with the time increases, the population of the open system diffuses
out and propagates in the finite bath (the periodic TLS chain); once
the population regathers back to the open system, the system dynamics
exhibits such a recurrence, and this process happens again and again,
which gives rise to the hierarchy recurrences.

We also study the dynamics of the \emph{total correlation entropy}
of the \noun{n}-body system, which sums up the entropy of all the
\noun{n }TLSs \citep{watanabe_information_1960,groisman_quantum_2005,zhou_irreducible_2008,anza_logarithmic_2020}.
It turns out the total correlation approximately exhibits a monotonic
increasing behavior, and the increasing curve becomes more and more
``smooth'' with the increase of the bath size. Thus, the total correlation
exhibits a quite similar behavior as the irreversible entropy increase
in the standard thermodynamics \citep{lebowitz_macroscopic_1993,li_production_2017,you_entropy_2018,li_correlation_2019}.
In contrast, the whole \noun{n}-body\noun{ }system always keeps a
constant due to the unitary evolution, and the entropy of each single
TLS increases and decreases from time to time. 

\vspace{0.8em}\noindent\textbf{Local relaxation: }We consider a chain
of \noun{n }TLSs. They have equal on-site energies ($\omega\ge0$),
and exchange energy with the nearest neighbors (interaction strengths
$g$): 
\begin{equation}
\hat{\mathcal{H}}=\sum_{n=0}^{\text{\textsc{n}}-1}\frac{1}{2}\omega\hat{\sigma}_{n}^{z}+g(\hat{\sigma}_{n}^{+}\hat{\sigma}_{n+1}^{-}+\hat{\sigma}_{n}^{-}\hat{\sigma}_{n+1}^{+}).\label{eq:H-0}
\end{equation}
Here $\hat{\sigma}_{n}^{+}:=(\hat{\sigma}_{n}^{-})^{\dagger}=|\mathtt{e}\rangle_{n}\langle\mathtt{g}|$,
$\hat{\sigma}_{n}^{z}:=|\mathtt{e}\rangle_{n}\langle\mathtt{e}|-|\mathtt{g}\rangle_{n}\langle\mathtt{g}|$,
and $|\mathtt{e}\rangle_{n}$, $|\mathtt{g}\rangle_{n}$ are the excited
and ground states of the $n$-th TLS. 

Here site-0 is regarded as an open ``\noun{system}'', while all
the other $(\text{\textsc{n}}-1)$ TLSs build up a finite ``\noun{bath}''.
Initially, the ``\noun{system}'' (site-0) starts from the excited
state as its initial state, and all the TLSs in the ``\noun{bath}''
start from the ground state. Thus effectively the ``\noun{bath}''
has a temperature $T\rightarrow0^{+}$. And now we study the dynamics
of the open ``\noun{system''.}

The Hamiltonian (\ref{eq:H-0}) is a quantum \emph{XX} model \citep{sachdev_quantum_2011},
and the dynamics of the whole chain is exactly solvable. Applying
the Jordan-Wigner transform, the Hamiltonian (\ref{eq:H-0}) becomes
a fermionic one, 
\begin{align}
\sigma_{n}^{z} & =2\hat{c}_{n}^{\dagger}\hat{c}_{n}-1,\quad\hat{\sigma}_{n}^{+}=\hat{c}_{n}^{\dagger}\,\prod_{i=0}^{n-1}(-\hat{\sigma}_{i}^{z}),\nonumber \\
\hat{\mathcal{H}} & =\sum_{n=0}^{\text{\textsc{n}}-1}\omega\hat{c}_{n}^{\dagger}\hat{c}_{n}+g(\hat{c}_{n}^{\dagger}\hat{c}_{n+1}+\hat{c}_{n+1}^{\dagger}\hat{c}_{n}).\label{eq:J-W-transform}
\end{align}
Under the periodic boundary condition, it can be further diagonalized
by the Fourier transform $\hat{c}_{n}=\sum_{k=0}^{\text{\textsc{n}}-1}\exp(i\frac{2\pi}{\text{\textsc{n}}}nk)\,\hat{b}_{k}\big/\sqrt{\text{\textsc{n}}}$,
which reads $\hat{\mathcal{H}}=\sum\varepsilon_{k}\hat{b}_{k}^{\dagger}\hat{b}_{k}$,
with the eigen mode energy $\varepsilon_{k}=\omega+2g\cos\frac{2\pi k}{\text{\textsc{n}}}$. 

The \noun{n}-body chain as a whole isolated system follows the unitary
evolution. From the above transformations, the above initial condition
gives $\langle\hat{c}_{0}^{\dagger}\hat{c}_{0}\rangle_{(t=0)}=1$,
and $\langle\hat{c}_{m}^{\dagger}\hat{c}_{n}\rangle_{(t=0)}=0$ for
the other $m,n$, and that gives the following dynamics 
\begin{align}
\langle\hat{c}_{m}^{\dagger}\hat{c}_{n}\rangle_{(t)} & =\sum_{k,q=0}^{\text{\textsc{n}}-1}\frac{1}{\text{\textsc{n}}}e^{i\frac{2\pi}{\text{\textsc{n}}}nq-i\frac{2\pi}{\text{\textsc{n}}}mk}\big\langle\hat{b}_{k}^{\dagger}(0)e^{i\varepsilon_{k}t}\cdot\hat{b}_{q}(0)e^{-i\varepsilon_{q}t}\big\rangle\nonumber \\
 & =\sum_{kq,xy}\frac{\langle\hat{c}_{x}^{\dagger}\hat{c}_{y}\rangle_{(0)}}{\text{\textsc{n}}^{2}}e^{i\frac{2\pi q}{\text{\textsc{n}}}(n-y)-i\frac{2\pi k}{\text{\textsc{n}}}(m-x)+i(\varepsilon_{k}-\varepsilon_{q})t}\nonumber \\
 & :=\big[\Phi_{m}^{(\text{\textsc{n}})}(2gt)\big]^{*}\,\Phi_{n}^{(\text{\textsc{n}})}(2gt),\label{eq:Cmn}
\end{align}
where we call $\Phi_{n}^{(\text{\textsc{n}})}(2gt:=\tau)$ as the
coherence function, and\footnote{\label{foot-Bessel}Utilizing $\exp[-i\tau\cos x]=\sum_{n=-\infty}^{\infty}(-i)^{n}J_{n}(\tau)e^{\pm inx}$.}\begin{subequations}
\begin{align}
\Phi_{n}^{(\text{\textsc{n}})}(\tau):= & \frac{1}{\text{\textsc{n}}}\sum_{k=0}^{\text{\textsc{n}}-1}\exp\big[-i\tau\,\cos\frac{2\pi k}{\text{\textsc{n}}}+i\frac{2\pi}{\text{\textsc{n}}}kn\big]\label{eq:Decay}\\
\stackrel{\text{\textsc{n}}\rightarrow\infty}{\longrightarrow} & \int_{0}^{2\pi}\frac{dx}{2\pi}\,e^{-i\tau\cos x+inx}=(-i)^{n}J_{n}(\tau).\label{eq:Jn(t)}
\end{align}
\end{subequations} In the thermodynamics limit $\text{\textsc{n}}\rightarrow\infty$,
$\Phi_{n}^{(\text{\textsc{n}})}(\tau)$ becomes the Bessel function
$J_{n}(\tau)$, which approaches zero when $\tau\rightarrow\infty$
\citep{hanggi_reaction-rate_1990,zwanzig_nonequilibrium_2001,cramer_exact_2008,flesch_probing_2008,eisert_quantum_2015}.

It can be seen from Eqs. (\ref{eq:J-W-transform}, \ref{eq:Cmn})
that, each site always keeps a diagonal density state $\hat{\rho}_{n}(t)=p_{n,\mathtt{e}}(t)|\mathtt{e}\rangle_{n}\langle\mathtt{e}|+p_{n,\mathtt{g}}(t)|\mathtt{g}\rangle_{n}\langle\mathtt{g}|$,
where $p_{n,\mathtt{e}}(t):=\langle\hat{\sigma}_{n}^{+}\hat{\sigma}_{n}^{-}\rangle_{(t)}=\langle\hat{c}_{n}^{\dagger}\hat{c}_{n}\rangle_{(t)}$
is the excited population of site-$n$. Therefore, if the ``\noun{bath}''
is infinitely large ($\text{\textsc{n}}\rightarrow\infty$), the ``\noun{system}''
would reach and stay at the ground state after long time relaxation,
namely, $p_{0,\mathtt{e}}(t)=|J_{0}(2gt)|^{2}\rightarrow0$ (here
the limit $\text{\textsc{n}}\rightarrow\infty$ is taken before $t\rightarrow\infty$).

\begin{figure}
\includegraphics[width=1\columnwidth]{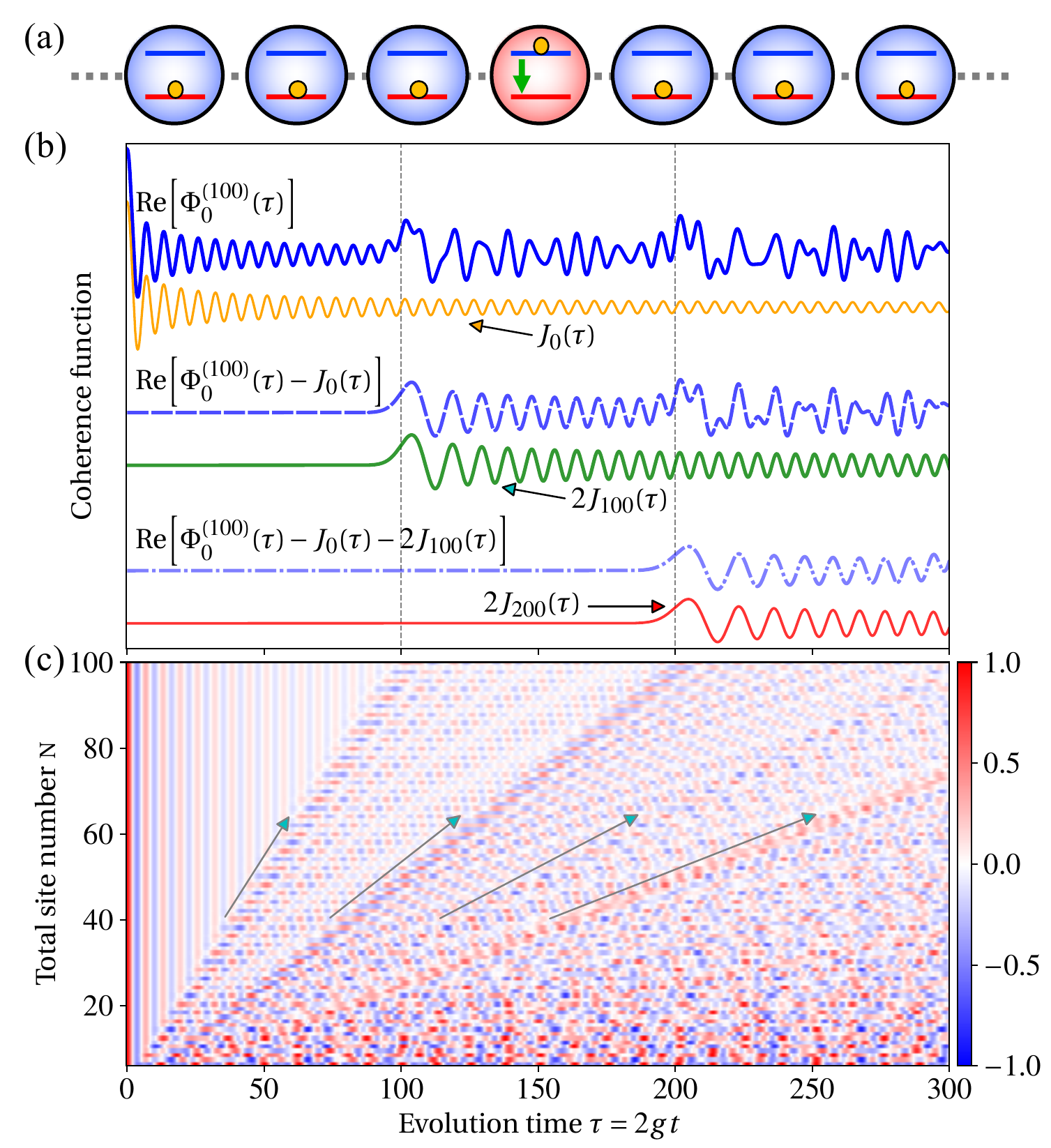}

\caption{(a) Demonstration for the \noun{n }TLSs and their initial states.
(b) The coherence function $\mathrm{Re}\big[\Phi_{0}^{(\text{\textsc{n}}=100)}(\tau)\big]$,
comparing with the Bessel functions. (c) The scaling behavior of $\mathrm{Re}\big[\Phi_{0}^{(\text{\textsc{n}})}(\tau)\big]$
with the site number \noun{n}.}

\label{fig-bessel}
\end{figure}

\vspace{0.8em}\noindent \textbf{Scaling behavior of recurrences:
}If the ``\noun{bath}'' is a finite one composed of $(\text{\textsc{n}}-1)$
TLSs, due to the finite-size effect, the above coherence functions
$\Phi_{n}^{(\text{\textsc{n}})}(\tau)$ exhibit a \emph{recurrence}
behavior\footnote{Precisely speaking, the recurrence behavior of
the open system here is different from the Poincar\'e recurrence
usually discussed in chaotic dynamics.}: within the time $0\le t\apprle t_{\text{rec}}:=\text{\textsc{n}}/2g$,
$\Phi_{n}^{(\text{\textsc{n}})}(\tau)$ fits the above Bessel function
(\ref{eq:Jn(t)}) quite closely and decays towards zero, but then
it shows a ``sudden bump'' and starts to look random after $t\apprge t_{\text{rec}}$
{[}see the solid blue line in Fig.\,\ref{fig-bessel}(b), $t_{\text{rec}}\equiv\text{\textsc{n}}/2g$
is the recurrence time{]} \citep{hanggi_reaction-rate_1990,zwanzig_nonequilibrium_2001,cramer_exact_2008,flesch_probing_2008,eisert_quantum_2015}.

Therefore, based on the local observation within a finite time smaller
than $t_{\text{rec}}$, we may conclude the open ``\noun{system}''
itself is relaxing towards a certain steady state, but indeed the
full \noun{n-}body state always keeps a pure state during the unitary
evolution. With the increase of the size \noun{n}, the recurrence
time $t_{\text{rec}}\equiv\text{\textsc{n}}/2g$ becomes larger and
larger, thus such a recurrence behavior does not show up in practice. 

In Fig.\,\ref{fig-bessel}(c), the scaling behavior of $\mathrm{Re}\big[\Phi_{0}^{(\text{\textsc{n}})}(\tau)\big]$
for different sizes \noun{n} is shown. Besides the above recurrence
appearing around $t\simeq t_{\text{rec}}\equiv\text{\textsc{n}}/2g$,
it is worth noting that some well-organized recurrence patterns also
appear in the region $t\apprge t_{\text{rec}}$. It can be seen similar
recurrences also appear periodically around $t\simeq\mathtt{q}\,t_{\text{rec}}$
for $\mathtt{q}=2,3,4,\dots$ {[}see the arrows in Fig.\,\ref{fig-bessel}(c){]}.
Moreover, each recurrence seems bringing in stronger randomness to
$\Phi_{0}^{(\text{\textsc{n}})}(\tau)$ than the previous one, which
forms a hierarchy structure, thus we call them \emph{hierarchy recurrences}.

We find that the appearance of such hierarchy recurrences can be explained
by the following expansion of $\Phi_{n}^{(\text{\textsc{n}})}(\tau)$
{[}Eq. (\ref{eq:Decay}){]}, that is$^{\ref{foot-Bessel}}$, 
\begin{align}
\Phi_{n}^{(\text{\textsc{n}})}(\tau)= & \frac{1}{\text{\textsc{n}}}\sum_{k=0}^{\text{\textsc{n}}-1}\big[\sum_{m=-\infty}^{\infty}(-i)^{m}J_{m}(\tau)e^{-im\cdot\frac{2\pi k}{\text{\textsc{n}}}}\big]\cdot e^{i\frac{2\pi}{\text{\textsc{n}}}kn}\nonumber \\
= & \sum_{m=-\infty}^{\infty}(-i)^{m}J_{m}(\tau)\cdot\frac{1}{\text{\textsc{n}}}\sum_{k=0}^{\text{\textsc{n}}-1}e^{i\frac{2\pi k}{\text{\textsc{n}}}(n-m)}\nonumber \\
= & \sum_{\mathtt{q}=-\infty}^{\infty}(-i)^{n+\mathtt{q}\text{\textsc{n}}}J_{n+\mathtt{q}\text{\textsc{n}}}(\tau).\label{eq:D(t)}
\end{align}
Here we used the relation $\sum_{k=0}^{\text{\textsc{n}}-1}e^{i\frac{2\pi k}{\text{\textsc{n}}}(n-m)}=\text{\textsc{n}}\delta_{n-m,\,\mathtt{q}\text{\textsc{n}}}$,
with $\mathtt{q}$ as an arbitrary integer.

For example, site-0 ($n=0$) gives a simple Bessel function series
{[}using $J_{-n}(\tau)=(-1)^{n}J_{n}(\tau)${]}\footnote{When $\text{\textsc{n}}\rightarrow\infty$,
the function series $\big\{\Phi_{0}^{(\text{\textsc{n}})}(\tau)\big\}$
converges pointwise to $J_{0}(\tau)$ but not uniformly.}
\begin{align}
\Phi_{0}^{(\text{\textsc{n}})}(\tau)= & J_{0}(\tau)+(-i)^{\text{\textsc{n}}}[1+(-1)^{\text{\textsc{n}}}]J_{\text{\textsc{n}}}(\tau)\nonumber \\
 & +(-i)^{2\text{\textsc{n}}}[1+(-1)^{2\text{\textsc{n}}}]J_{2\text{\textsc{n}}}(\tau)+\dots
\end{align}
For a large \noun{n}, the Bessel function $J_{\text{\textsc{n}}}(\tau)\simeq0$
in the area $0\le\tau\apprle\text{\textsc{n}}$, and starts to exhibit
significant oscillations after $\tau\simeq\text{\textsc{n}}$ {[}see
$J_{100}(\tau)$ in Fig.\,\ref{fig-bessel}(b){]}. Therefore, in
the above expansion of $\Phi_{0}^{(\text{\textsc{n}})}(\tau)$, each
term $J_{\mathtt{q}\text{\textsc{n}}}(\tau)$ contributes a ``sudden
bump'' around $\tau\simeq\mathtt{q}\text{\textsc{n}}$, and this
is just why the above recurrences appear around $t\simeq\mathtt{q}\,t_{\text{rec}}$
($\mathtt{q}=1,2,3,\dots$).

\begin{figure}
\includegraphics[width=1\columnwidth]{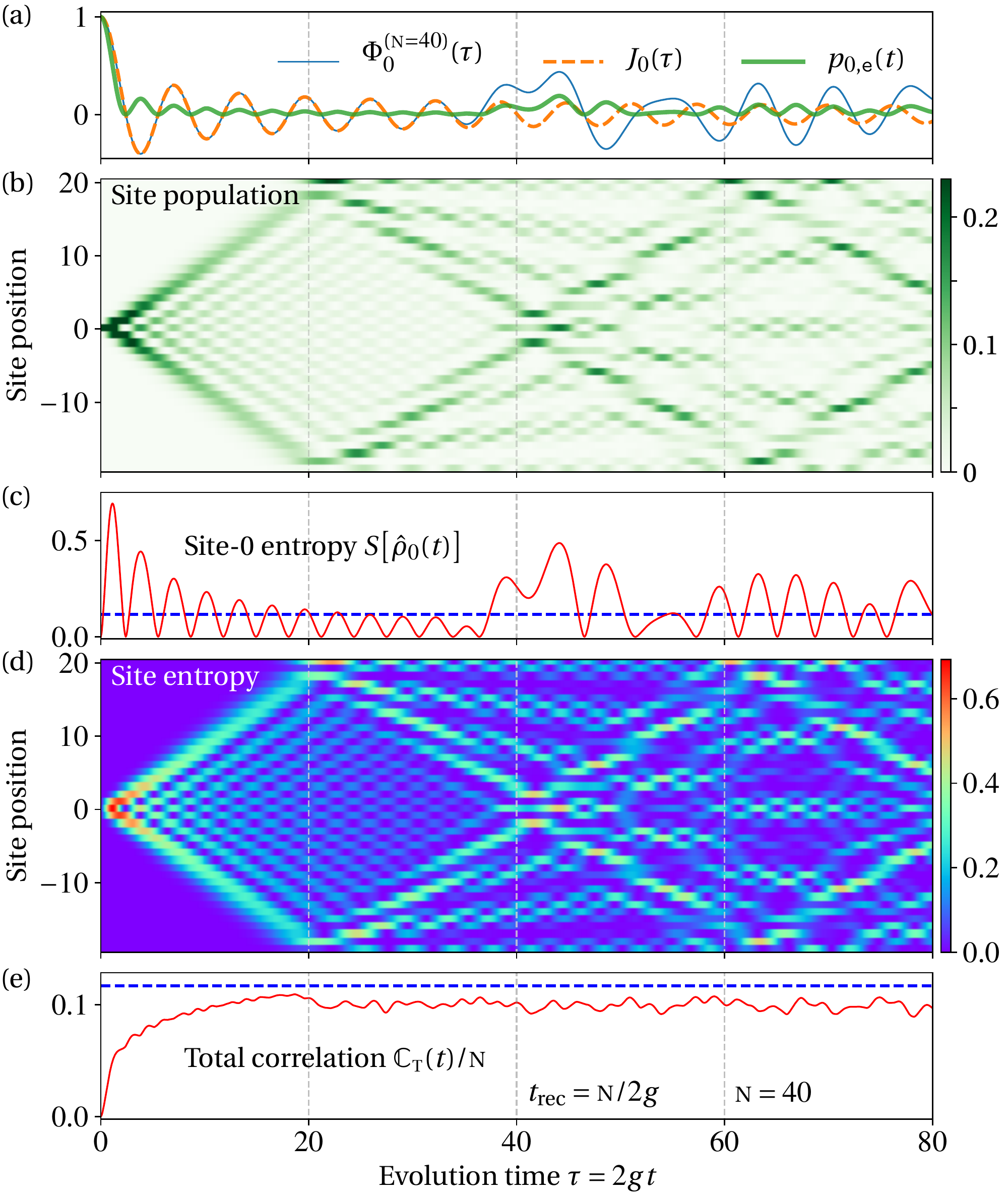}

\caption{(a) The coherence function $\Phi_{0}^{(\text{\textsc{n}})}(\tau)$
and site-0 population $p_{0,\mathtt{e}}(t)$. (b) The population evolution
of each TLS. (c) The entropy dynamics of site-0. (d) The entropy of
each TLS. (e) The total correlation $\mathbb{C}_{\text{\textsc{t}}}(t)/\text{\textsc{n}}$.
The dashed blue lines in (c, e) are $\mathbb{C}_{\text{max}}^{(\text{\textsc{n}})}/\text{\textsc{n}}$.
Here the site number is $\text{\textsc{n}}=40$.}

\label{fig-reflect}
\end{figure}

\vspace{0.8em}\noindent \textbf{Population propagation: }The population
dynamics of all the \noun{n} TLSs is shown in Fig. \ref{fig-reflect}(b),
i.e., $p_{n,\mathtt{e}}(t)=\langle\hat{\sigma}_{n}^{+}\hat{\sigma}_{n}^{-}\rangle_{(t)}=\big|\Phi_{n}^{(\text{\textsc{n}})}(2gt\equiv\tau)\big|^{2}$,
and a propagation pattern is clearly seen. Initially, the population
distribution of the \noun{n} TLSs forms a ``cusp'' around site-0
{[}$p_{0,\mathtt{e}}(t)=1$, and $p_{n,\mathtt{e}}(t)=0$ for $n\neq0${]}.
Within the time $t\apprle t_{\text{rec}}$, the initial population
``cusp'' on site-0 propagates towards the two directions of the
periodic chain, and the propagation ``speed'' is almost a constant
\citep{lieb_finite_1972,ganahl_observation_2012}. This constant speed
also can be seen from the leading terms of $p_{n,\mathtt{e}}(t)=\big|\Phi_{n}^{(\text{\textsc{n}})}(2gt)\big|^{2}\simeq\big|J_{|n|}(2gt)\big|^{2}+\dots$
{[}for $-\text{\textsc{n}}/2<n<\text{\textsc{n}}/2$, see Eq. (\ref{eq:D(t)}){]}:
the leading Bessel function indicates the first ``sudden bump''
of site-$n$ appears around $t\simeq|n|/2g$, which linearly depends
on the distance $|n|$ to site-0 {[}here site-$(-n)$ and site-$(\text{\textsc{n}}-n)$
are the same one due to the periodic boundary condition{]}.

The two-side propagations would meet each other at the periodic boundaries
at $n\sim\pm\text{\textsc{n}}/2$, and then regathers back to site-0
again. Notice that this is just the moment that $\Phi_{0}^{(\text{\textsc{n}})}(2gt)$
exhibits its first recurrence ($t\simeq t_{\text{rec}}\equiv\text{\textsc{n}}/2g$,
see the dashed vertical lines in Fig. \ref{fig-reflect}). The propagation
regathered back would be superposed with the original one, and that
makes the system dynamics appear more random. Clearly, since such
propagation and regathering happens again and again, the ``\noun{system}''
(site-0) experiences the above hierarchy recurrences periodically
around $t\simeq\mathtt{q}\,t_{\text{rec}}$.

\vspace{0.8em}\noindent \textbf{Total correlation entropy: }Now we
consider the entropy dynamics in this system. The \noun{n}-body chain
as a whole isolated system follows the unitary evolution, thus its
von Neumann entropy $S[\hat{\boldsymbol{\rho}}(t)]=-\mathrm{tr}[\hat{\boldsymbol{\rho}}\ln\hat{\boldsymbol{\rho}}]$
always keeps zero as the initial state. Since initially the ``\noun{bath}''
has a zero temperature $T\rightarrow0^{+}$, the thermal entropy $\text{\dj}Q/T$
in the standard thermodynamics cannot be used here either \citep{santos_wigner_2017}. 

Notice that, in practical observations, indeed the full \noun{n}-body
state is usually not directly accessible for local measurements, and
it is the few-body observables that can be directly measured \citep{swendsen_explaining_2008,li_correlation_2019,strasberg_entropy_2019}.
Therefore, here we consider the dynamics of the \emph{total correlation
entropy }of the \noun{n}-body state $\hat{\boldsymbol{\rho}}(t)$,
that is \citep{watanabe_information_1960,groisman_quantum_2005,zhou_irreducible_2008,anza_logarithmic_2020},
\begin{equation}
\mathbb{C}_{\text{\textsc{t}}}[\hat{\boldsymbol{\rho}}(t)]:=\sum_{n=0}^{\text{\textsc{n}}-1}S[\hat{\rho}_{n}(t)]-S[\hat{\boldsymbol{\rho}}(t)],\label{eq:CorrEnt}
\end{equation}
where $\hat{\rho}_{n}$ are the reduced 1-body states. For bipartite
systems ($\text{\textsc{n}}=2$), it just returns the mutual information
\citep{nielsen_quantum_2000,li_production_2017,li_correlation_2019,you_entropy_2018}.

\begin{figure}
\includegraphics[width=1\columnwidth]{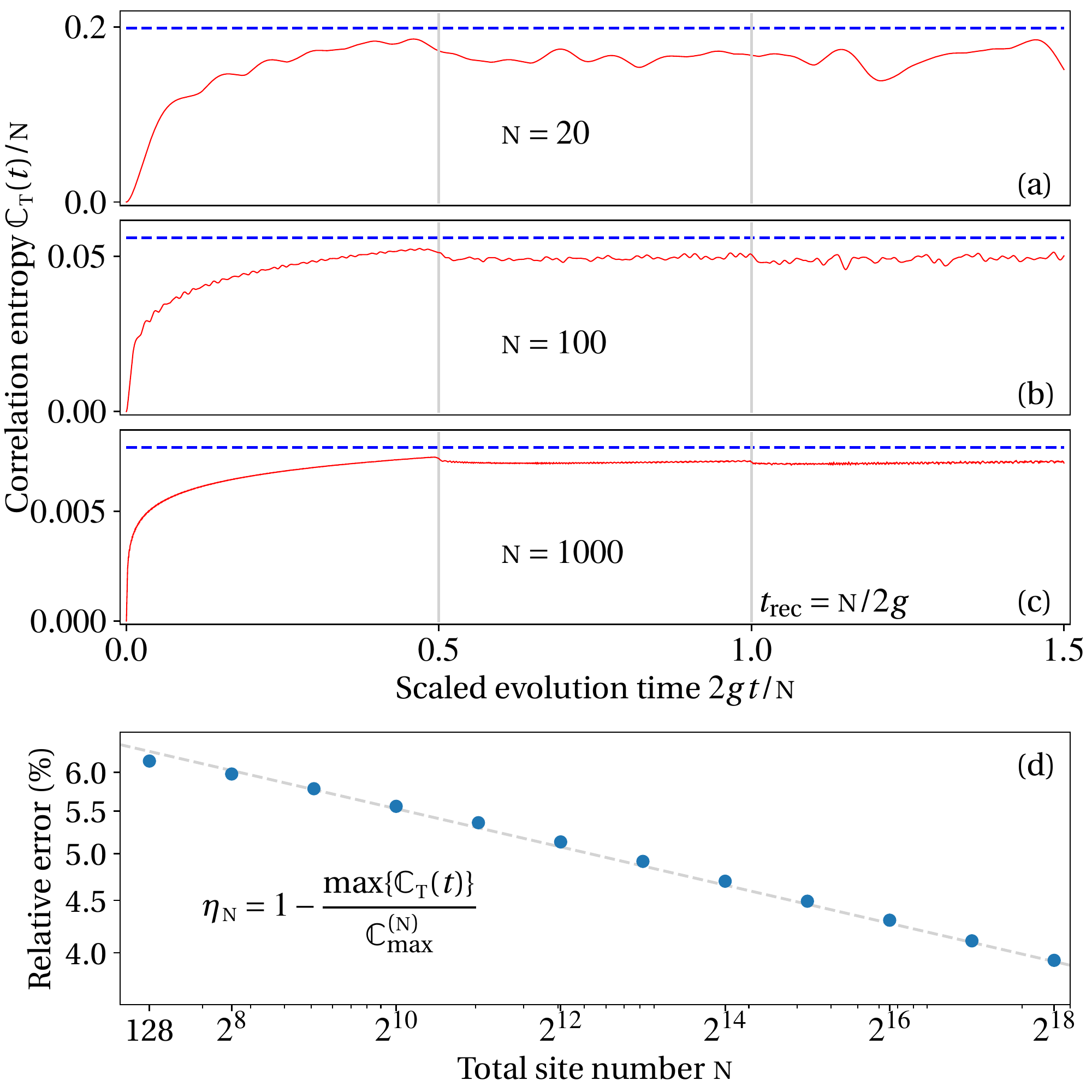}

\caption{(a-c) Evolution of the total correlation $\mathbb{C}_{\text{\textsc{t}}}(t)/\text{\textsc{n}}$
with the scaled time $2gt/\text{\textsc{n}}$ for $\text{\textsc{n}}=20,100,1000$.
The blue dashed lines are the correlation maximum $\mathbb{C}_{\text{max}}^{(\text{\textsc{n}})}/\text{\textsc{n}}$.
(d) The relative error $\eta_{\text{\textsc{n}}}$ between $\text{max}\big\{\mathbb{C}_{\text{\textsc{t}}}(t)\big\}$
(around $t\simeq t_{\text{rec}}/2$) and the maximum $\mathbb{C}_{\text{max}}^{(\text{\textsc{n}})}$
decreases with the site number \noun{n}.}

\label{fig-entropy}
\end{figure}

It turns out, the entropy of each single TLS is increasing and decreasing
from time to time, and they also have the above recurrence behavior
{[}Fig. \ref{fig-entropy}(c, d){]}. In contrast, their summation
as the total correlation $\mathbb{C}_{\text{\textsc{t}}}(t)$ approximately
exhibits a monotonic increasing behavior {[}except still carrying
small fluctuations, see Fig. \ref{fig-reflect}(e){]}. Moreover, with
the increase of the chain size \noun{n}, the increasing curve of $\mathbb{C}_{\text{\textsc{t}}}(t)$
appears more and more ``smooth'' {[}Fig. \ref{fig-entropy}(a-c){]}.
Clearly, this is quite similar with the behavior of the irreversible
entropy production during the relaxation process in the standard thermodynamics
\citep{spohn_entropy_1978,de_groot_non-equilibrium_1962,kondepudi_modern_2014,nicolis_self-organization_1977}.

Now we consider the correlation maximum that $\mathbb{C}_{\text{\textsc{t}}}(t)$
might achieve \citep{jaynes_information_1957}. With the help of Lagrangian
multipliers, under the constraints (1) $p_{n,\mathtt{g}}+p_{n,\mathtt{e}}=1$
(probability normalization) and (2) $\sum_{n}p_{n,\mathtt{e}}=1$
(excitation number conservation), the maximum of $\mathbb{C}_{\text{\textsc{t}}}\big[\{p_{n,\mathtt{e}(\mathtt{g})}\}\big]=\sum_{n}-p_{n,\mathtt{e}}\ln p_{n,\mathtt{e}}-p_{n,\mathtt{g}}\ln p_{n,\mathtt{g}}$
is obtained as 
\begin{equation}
\mathbb{C}_{\text{max}}^{(\text{\textsc{n}})}=\text{\textsc{n}}\ln\frac{\text{\textsc{n}}}{\text{\textsc{n}}-1}+\ln(\text{\textsc{n}}-1).\label{eq:maxCT}
\end{equation}
The maximum is achieved when all the \noun{n }TLSs have the same populations
$\tilde{p}_{n,\mathtt{e}}=1/\text{\textsc{n}}$. 

Under the scaled time $2gt/\text{\textsc{n}}$, the correlation evolutions
$\mathbb{C}_{\text{\textsc{t}}}(t)/\text{\textsc{n}}$ for different
sizes \noun{n }appear quite similar to each other {[}Fig.\,\ref{fig-entropy}(a-c){]}.
They all approach their upper bound $\mathbb{C}_{\text{max}}^{(\text{\textsc{n}})}/\text{\textsc{n}}$
closely, and come across a ``sudden bump'' around half of the recurrence
time $t\simeq t_{\text{rec}}/2$ (indeed this is just the moment the
two-side propagations meet each other at $n\sim\pm\text{\textsc{n}}/2$).

We denote $\eta_{\text{\textsc{n}}}:=1-\text{max}\{\mathbb{C}_{\text{\textsc{t}}}(t)\}\big/\mathbb{C}_{\text{max}}^{(\text{\textsc{n}})}$
as the relative error between $\text{max}\{\mathbb{C}_{\text{\textsc{t}}}(t)\}$
and the correlation maximum $\mathbb{C}_{\text{max}}^{(\text{\textsc{n}})}$.
With the increase of the size \noun{n}, the error $\eta_{\text{\textsc{n}}}$
decays slowly towards zero {[}approximately $\eta_{\text{\textsc{n}}}\propto\text{\textsc{n}}^{-\alpha}$
with $\alpha\simeq0.062$, see Fig.\,\ref{fig-entropy}(d){]}. In
the thermodynamic limit $\text{\textsc{n}}\rightarrow\infty$, we
may expect $\mathbb{C}_{\text{\textsc{t}}}(t)$ could reach the maximum
$\mathbb{C}_{\text{max}}^{(\text{\textsc{n}})}$. 

In this sense, the above correlation maximization effectively gives
a \emph{pseudo-equilibrium state} $\tilde{\boldsymbol{\rho}}_{\text{eq}}\equiv{\textstyle \bigotimes_{n}}\,\tilde{\varrho}_{n}$,
where $\tilde{\varrho}_{n}:=\frac{1}{\text{\textsc{n}}}|\mathtt{e}\rangle_{n}\langle\mathtt{e}|+(1-\frac{1}{\text{\textsc{n}}})|\mathtt{g}\rangle_{n}\langle\mathtt{g}|$,
and the whole \noun{n}-body state $\hat{\boldsymbol{\rho}}(t)$ ``looks''
like approaching this pseudo-equilibrium state during the unitary
evolution \citep{cramer_exact_2008}. But we emphasize indeed $\hat{\boldsymbol{\rho}}(t)$
and $\hat{\rho}_{n}(t)$ never have any steady states when $t\rightarrow\infty$,
and $\hat{\boldsymbol{\rho}}(t)$ always keeps a pure state.

The increasing rate of the above total correlation (\ref{eq:CorrEnt})
also can be rewritten in the form of relative entropy \citep{spohn_entropy_1978,esposito_entropy_2010,manzano_entropy_2016,ptaszynski_entropy_2019}
\begin{equation}
\partial_{t}\mathbb{C}_{\text{\textsc{t}}}(t)=\partial_{t}D[\hat{\boldsymbol{\rho}}(t)\parallel{\textstyle \bigotimes_{n}}\,\hat{\rho}_{n}(t)],\label{eq:RelativeEntr}
\end{equation}
where $D[\rho\Vert\varrho]=\mathrm{tr}[\rho(\ln\rho-\ln\varrho)]$
is the relative entropy. Approximately, the reference state ${\textstyle \bigotimes_{n}}\,\hat{\rho}_{n}(t)$
here can be replaced by the pseudo-equilibrium state $\tilde{\boldsymbol{\rho}}_{\text{eq}}\equiv{\textstyle \bigotimes_{n}}\,\tilde{\varrho}_{n}$. 

We emphasize that the pseudo-equilibrium state $\tilde{\boldsymbol{\rho}}_{\text{eq}}$
here is determined by the above correlation maximization, but irrelevant
with the on-site energy $\omega$ and the interaction strength $g$,
thus it is different from the canonical state like $\hat{\boldsymbol{\rho}}_{\text{th}}\sim\exp[-\hat{\mathcal{H}}/T]$.
If the on-site energy $\omega\le0$, all the above results for the
system dynamics still remains the same.

Moreover, when $\omega<0$, the status of $|\mathtt{g}\rangle_{n}$
and $|\mathtt{e}\rangle_{n}$ are indeed reversed: initially, the
open ``\noun{system}'' starts from the ground state while the TLSs
in the ``\noun{bath}'' start from the excited state. Therefore,
effectively the ``\noun{bath}'' has a negative temperature $T\rightarrow0^{-}$
\citep{landau_statistical_1980,ramsey_thermodynamics_1956}. Notice
that all the above results of the total correlation entropy still
applies in this situation.

\vspace{0.8em}\noindent \textbf{Summary: }In this paper, we study
the local relaxation process of an open system contacted with a finite
bath. We find that, due to the finite-size effect of the bath, the
local relaxation of the open system exhibits hierarchy recurrences
periodically, which makes the system dynamics appear more and more
random. Essentially, that is because the energy diffuses out of the
open system regathers back from the finite bath again and again. During
the unitary evolution, the open system and the bath as a whole isolate
system keeps a constant entropy, and the entropy of each single TLS
increases and decreases from time to time, while the total correlation
entropy approximately exhibits a monotonic increasing behavior, which
is similar as the irreversible entropy increase in the standard thermodynamics
\citep{lebowitz_macroscopic_1993,li_production_2017,you_entropy_2018,li_correlation_2019}.
We emphasize throughout the above discussions there is no average
on time or any random configurations. The quantum \emph{XX} model
here could be realized in many physical systems, such as optical lattices
\citep{trotzky_probing_2012}, superconducting circuits \citep{ye_propagation_2019,orell_probing_2019},
and ion trap arrays \citep{smith_many-body_2016,xiong_experimental_2018}.

\vspace{0.5em} \emph{Acknowledgment} -- S.-W. Li appreciates for
the helpful discussions with J. Chen, B. Garraway, N. Wu, D. Z. Xu,
Y.-N. You. This study is supported by NSF of China (Grant No.\,11905007),
Beijing Institute of Technology Research Fund Program for Young Scholars.

\bibliographystyle{apsrev4-2}
\bibliography{Refs}
 
\end{document}